\documentclass{ws-procs9x6}
\usepackage{graphicx}
\usepackage{fancyhdr}
\usepackage{subfigure}
\usepackage{epstopdf}

\begin{document}

\title{CR electrons and positrons: what we have learned in the
latest three years and future perspectives}

\author{Daniele Gaggero, Dario Grasso}

\address{Department of Physics, Pisa University,\\
Largo B. Pontecorvo 3, 56127 Pisa Italy\\
$^*$E-mail: {\tt daniele.gaggero@pi.infn.it}\\
}

\begin{abstract}
After the PAMELA finding of an increasing positron fraction above 10 GeV,  the experimental evidence of the presence of a new electron and positron spectral component in the cosmic ray zoo has been recently confirmed by Fermi-LAT. We show as a simple phenomenological model which assumes the presence of an electron and positron extra component peaked at $\sim 1~{\rm TeV}$ allows a consistent description of all available data sets. We then describe the most relevant astrophysical uncertainties which still prevent to determine $e^\pm$ source properties from those data and the perspectives of forthcoming experiments.  
\end{abstract}


\bodymatter

\section{Introdution}\label{Introduction}

Recent experimental results raised a wide interest about the origin and the propagation of the leptonic component of the cosmic radiation.  

Among the most striking of those results, there is the observation performed by the PAMELA satellite experiment that the positron to electron fraction $e^+/ (e^- + e^+)$ rises with energy from 10 up to 100 GeV at least (Adriani {\it et al.} 2008 \cite{Adriani_Nature_2008}). This appeared in contrast with the predictions of the standard cosmic ray scenario and could therefore be interpreted as the smoking gun of new physics, unless a very soft electron spectrum was assumed.  

The significance of this anomaly increased when the Fermi-LAT space observatory measured the $e^- + e^+$ spectrum in the 7 GeV - 1 TeV energy range with unprecedented accuracy and found it to be compatible with a power-law with index $\gamma(e^\pm) = - 3.08 \pm 0.05 $ (Abdo {\it et al.} 2009 \cite{Fermi_el_2009}, Ackermann {\it et al.} 2010 \cite{Fermi_el_2010}); this slope is significantly harder than what estimated on the basis of previous measurements: the hypothesis of a steep spectrum was therefore excluded.  

More recently, the same collaboration provided a further, and stronger, evidence of the positron anomaly by providing direct measurement of the absolute $e^+$ and $e^-$
spectra, and of their fraction, between 20 and ~200 GeV using the Earth magnetic field. A steady rising of the positron fraction was observed by this experiment up to that energy in agreement with that found by PAMELA.  In the same energy range, the $e^-$ spectrum was fitted with a power-law with index $\gamma(e^-) = - 3.19 \pm 0.07$ which is in agreement with what recently measured by PAMELA between 1 and  625 GeV (Adriani {\it et al.} 2011 \cite{pamela_el_2011}).  Most importantly, Fermi-LAT measured, for the first time, the $e^+$ spectrum in the 20 - 200 GeV energy interval and showed it is fitted by a power-law with index  $\gamma(e^+) =  - 2.77  \pm 0.14$.  

We will show in the following paragraph how all those measurements rule out the standard scenario in which the bulk of electrons reaching the Earth in the GeV - TeV energy range are originated by Supernova Remnants (SNRs) and only a small fraction of secondary positrons and electrons comes from the interaction of CR nuclei with the interstellar medium (ISM). 
Then we will see how the alternative scenario in which the presence of electron + positron component peaked at  $\sim 1$ TeV is invoked allows a consistent description of all the available data sets.   
Finally we will discuss to which extent astrophysical and particle physics uncertainties still affect our modeling of cosmic ray leptons origin and propagation and how forthcoming measurements are expected to reduce those uncertainties.

\section{The necessity of a primary extra-component}\label{sec:extra_comp}

%
%
%
%
%
%

After the release of Fermi-LAT $e^- + e^+$ spectrum, it was clearly pointed out in several papers (see e.g. Grasso {\it et al.} 2009 \cite{Fermi_el_interpretation} and Di Bernardo {\it et al. 2011} \cite{electrons_final}) that both Fermi-LAT and PAMELA measurements described in the Introduction are in contrast with a standard single-component scenario in which positrons are the secondary products of the nuclear component of cosmic rays (CRs) interacting with the interstellar medium (ISM).


\begin{figure}[h]
	\setlength{\unitlength}{1mm}
     \begin{center}
      \includegraphics[width=8.cm]{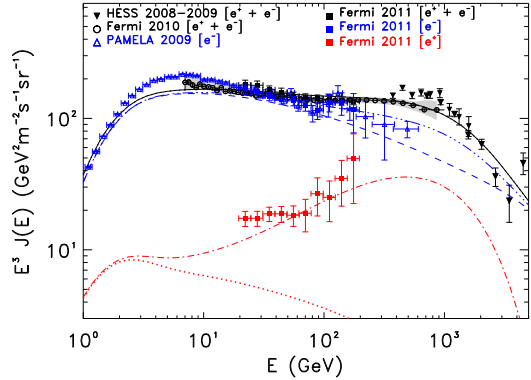} 
     \end{center}
     \caption{\footnotesize \it Fermi-LAT and PAMELA data on electrons + positrons and electrons are compared to a double component phenomenological model. The absolute positron spectrum is compared to a single and double component phenomenological model.  
     {\bf Red dotted line}: $e^+$ in single-component scenario.
     {\bf Red dot-dashed line}: $e^+$ in double-component scenario.
     {\bf Blue triple dotted-dashed line}, {\bf black solid line}: $e^-$ and $e^- + e^+$ in double-component scenario.
     {\bf Blue dashed line}: $e^-$ diffuse background in double-component scenario.
     The Kolmogorov diffusion setup is adopted. }
     \label{fig:elepos}
\end{figure}

The main problems encountered by this kind of models can be summarized as follows.

\begin{itemize}
\item As explained many times (see e.g. Serpico 2011 \cite{Serpico2011} for a recent review), they cannot reproduce the rising positron-to-electron ratio measured by PAMELA and recently confirmed by Fermi-LAT;
\item They are unable to reproduce all the features revealed by Fermi-LAT in the CRE spectrum, in particular the flattening observed at around 20 GeV and the softening at $\sim 500$ GeV. In fact, if such models are normalized against data in the  20 - 100 GeV energy range, where systematical and theoretical uncertainties are the smallest, they clearly fail to match CRE Fermi-LAT and PAMELA $e^-$ data outside that range. A different normalization results in even worse fits. 

\end{itemize} 

With the release of the $e^-$ and $e^+$ separate spectra by the Fermi-LAT collaboration the problems with the single component scenario became even worse. 
In fact, the $e^+$ spectrum (Fig. \ref{fig:elepos}) is clearly inconsistent with the predictions of a single component scenario computed with {\tt DRAGON} numerical diffusion package (and similar results are obtained with {\tt GALPROP}). 
Even without considering numerical models, the simple consideration that the reported positron spectral slope is $-2.77 \pm 0.14$ reveals how these data are incompatible with a purely secondary origin from proton spallation on interstellar gas: the source slope should be the same as the proton spectrum, i.e. $\simeq -2.75$  (Adriani {\it et al.} 2011 \cite{Pamela_p_2011}) and no room is then left for the unavoidable steepening due to energy-dependent diffusion and energy losses.



\begin{figure}[h]
	\setlength{\unitlength}{1mm}
     \begin{center}
      \includegraphics[width=8.cm]{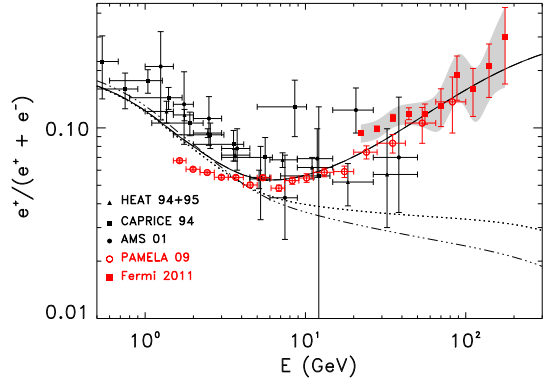} 
     \end{center}
     \caption{\footnotesize \it Fermi-LAT and PAMELA data on the positron ratio are compared to a single and double component phenomenological model.
     {\bf Dot-dashed line}: positron ratio in single-component scenario.
     {\bf Dotted line}: positron ratio in double-component scenario due to conventional secondary positron production.
     {\bf Solid line}: positron ratio in double-component scenario {\it including} extra-component.
     The progagation setup and modulation potential are the same of Fig. 1. The solar modulation potential is taken $\Phi = 550$ MV in all figures of this paper.}
     \label{fig:posratio}
\end{figure}

A double component scenario is the most straightforward solution to these problems. 

The idea dates back to the pioneering work by F. Aharonian and A. Atoyan 1995 \cite{AharonianAtoyan1995} and was extensively studied after the release of ATIC and PAMELA data in 2008 (see e.g. Hooper {\it et al. 2009} \cite{Hooper_2009} and Profumo 2008 \cite{Profumo_2008}).

More recently, we contributed to several papers in which it was shown that a consistent interpretation of the $e^+ + e^-$ spectrum measured by Fermi-LAT and the PAMELA positron fraction can be naturally obtained in that framework (Grasso {\it et al.} 2009 \cite{Fermi_el_interpretation}, Ackermann {\it et al.} 2010 \cite{Fermi_el_2010}, Di Bernardo {\it et al.}  2011 \cite{electrons_final}). 

For example, in Fig. \ref{fig:elepos} and Fig. \ref{fig:posratio} we show that the double component model proposed in Ackermann {\it et al.} 2010\cite{Fermi_el_2010} reproduces the data mentioned above and also the $e^+$ and $e^-$ separate spectra, and their ratio, recently released by the Fermi-LAT collaboration and not yet available at the time.
The model represented in those figures assumes a propagation setup characterized by a cylindrical diffusive halo with half-thikness of 4 kpc; a diffusion coefficient scaling with rigidity like $\rho^{1/3}$ (corresponding to a Kolmogorov-like diffusion within the quasi-linear approximation) and a relatively strong reacceleration  (the Alfv\'en velocity is $v_A = 30~{\rm km s^{-1}}$).  Solar modulation is treated here as
charge independent in the force field approximation by fixing the modulation potential $\Phi$ against proton data taken in the same solar phase.  
In that model, the standard $e^-$ primary component is tuned to fit Fermi-LAT data at low energy in the presence of the extra-component becoming dominant at higher energies;  the injection slope for the primary electron component is set to $-2.70$ above 2 GeV, while under that energy a slope of $-1.6$ is adopted, in accord with recent constraints from the synchrotron spectra (see Jaffe et al. 2011 \cite{JaffeStrong2011}).
The extra component, instead, originates from a primary source of electron+positron pairs; it has an injection spectrum modelled in a simple way as a power-law with index $-1.5$ plus an exponential cutoff at $1.2$ TeV; the spatial distribution of this source is the same as the standard one and the propagation parameters are also the same; the normalization is tuned so that Fermi-LAT and PAMELA data at high energy are matched by the sum of standard + extra component. 
Both components are computed with {\tt DRAGON} (even if it was checked that the same result can be obtained with {\tt GALPROP}).

An issue remains open about the origin of the discrepancy between the prediction of this, or similar, models and the positron fraction measured by PAMELA below 10 GeV. 
In the next section we will show as that discrepancy may be interpreted as the consequence of an incorrect choice of the propagation setup and discuss other uncertainties which can affect 
the electron and positron spectra in that low energy range.  


\section{LOW ENERGY. Impact of astrophysical uncertainties}\label{sec:low_energy}

%

\begin{figure}[h]
	\setlength{\unitlength}{1mm}
     \begin{center}
      \includegraphics[width=8.cm]{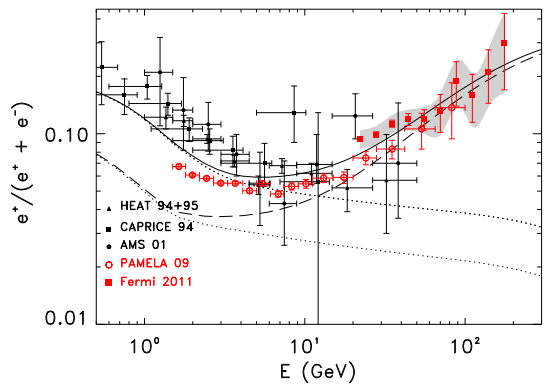}
 	 \end{center}
     \caption{\footnotesize \it Effect of changing the diffusion halo height.
     Solid line: h = 1 kpc; dashed: h = 10 kpc.}
     \label{fig:doublecomponent_haloheight}
\end{figure}

\begin{figure}[h]
	\setlength{\unitlength}{1mm}
     \begin{center}
      \includegraphics[width=8.cm]{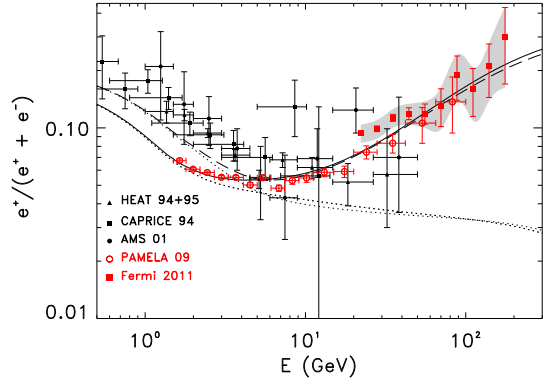}
 	 \end{center}
     \caption{\footnotesize \it Effect of the diffusion setup.
     Solid line: KRA; dashed line: KOL.}
     \label{fig:doublecomponent_setup}
\end{figure}

Cosmic ray electrons and positrons, either belonging to the standard or the extra component, propagate in the Galaxy undergoing several physical processes: diffusion, reacceleration, energy losses. Such complex motion is effectively described by a well-known diffusion-loss equation (Berezinskii {\it et al.} 1990 \cite{Berezinsky_book}). In this equation several free parameters are involved: the height of the halo in which the propagation takes place, the normalization and energy dependence of the diffusion coefficient (the latter parametrized by the parameter $\delta$), the Alfv\'en velocity that influences the effectiveness of reacceleration; moreover, several astrophysical inputs need to be considered: the injection spectrum, the spatial distribution of the source term, the interstellar radiation field, the gas distribution. 

The free parameters that appear in the diffusion-loss equation are constrained by some CR observables such as Boron-to-Carbon (B/C) or antiproton-to-proton ratio; different {\it diffusion setups} exist in the literature, obtained through comparison of experimental data with the prediction of semi-analytical codes (Maurin {\it et al.} 2001 \cite{Maurin2001}, Donato {\it et al.} 2004 \cite{Donato2004}) or numerical packages such as {\tt DRAGON} or {\tt GALPROP} (see e.g. Di Bernardo {\it et al.} 2010 \cite{anstat_2010} for {\tt DRAGON}-related models and Trotta {\it et al.} 2011 \cite{Trotta2011} for a {\tt GALPROP}-based analysis).

The uncertainties related to the diffusion model and to the astrophysical inputs were discussed in the latest years in several papers making use of semi-analytic codes (e.g. Delahaye {\it et al.} 2010 \cite{Delahaye_2010}). In the following we will briefly analyse the impact of these uncertainties adopting the {\tt DRAGON} code.

One of the most relevant parameter is the halo height. According to the analytical computations by Bulanov and Dogel \cite{BulanovDogel1974}, while at low energy the electrons (or positrons) are distributed throughout all the diffusion halo, as the energy increases the electrons occupy a smaller and smaller fraction of the halo due to energy losses.  
This is relevant especially for the secondary positron spectrum. In fact, since their injection power is determined by the CR nuclei density in the Galactic disk, a thicker halo results in a larger dilution of their density in the halo hence a in smaller flux on the Earth.   
Numerical computations confirm the expectation of this heuristic argument as shown in Fig. \ref{fig:doublecomponent_haloheight}. From the plot it is also evident that large halo heights are disfavoured by the data.

Even fixing the height of the diffusion halo, the choice of the diffusion setup can also affect the low energy spectra of CR leptons. 
This is evident from Fig. \ref{fig:doublecomponent_setup} where we compare the predictions of two different models which both reproduce nuclear CR data:

\begin{itemize}
\item a Kraichnan-like diffusion setup with $\delta = 0.5$ and moderate reacceleration (that was pointed out as the preferred one in a {\tt DRAGON}-based maximum likelihood analysis with focus on both B/C and antiproton high energy data\cite{anstat_2010})

\item a Kolmogorov-like diffusion setup with $\delta = 0.33$ and high reacceleration (that was pointed out as the preferred one in a {\tt GALPROP}-based maximum likelihood analysis with focus on B/C data\cite{Trotta2011})
\end{itemize}

It is clear from that plot that the Kraichnan-like setups allows a better fit of low-energy positron ratio measured by PAMELA; this consideration, together with several other facts (high reacceleration models do not permit a good fit of antiproton data and cannot reproduce the spectrum of the synchrotron emission of the Galaxy), led us to conclude that models with strong reacceleration are disfavoured. 

\section{High energy uncertainties and the nature of the extra-component}


In the double component scenario discussed in Sec. \ref{sec:extra_comp}, the positron spectrum above $\sim 10$ GeV is dominated by the primary extra component. 
The nature of its source is one of the hottest matter of debate in the CR physics. 

Galactic pulsars were suggested as natural source candidates of a primary CR positron component well before PAMELA results (Aharonian and Atoyan, 1995 \cite{AharonianAtoyan1995}.) 
More recently, it was noticed that a single, nearby, pulsar (such as Monogem or Geminga) could explain the positrons fraction excess found by PAMELA (Hooper {\it et al.} 2009 \cite{Hooper_2009}). 

In the Fermi-LAT era, we showed (Grasso {\it et al. 2009} \cite{Fermi_el_interpretation} and Di Bernardo {\it et al.} 2010 \cite{electrons_final}) that also the $e^+ + e^-$ measured by that experiment can consistently be explained in the same terms: if one considers the observed nearby pulsars within 2 kpc and assumes that a relevant fraction of their rotational energy is transferred into $e^+ + e^-$ pairs ($\simeq 30$\%), under reasonable assumpions on the injection spectrum and cutoff it is possible to reproduce all existing data. 
In the cosmic ray channel, this scenario has two possible testable consequences:
 
\begin{itemize}
\item the detection of a CR electron anisotropy towards the most relevant sources (in our analysis, Monogem and Geminga \cite{Hooper_2009}); 
\item the presence of some bumpiness in the $e^-$ and $e^+$ spectra in the TeV region due to the contribution of several pulsars. 
\end{itemize}

Those two signatures are somehow complementary: if a single pulsar give the dominant contribution to the extra component a large anisotropy and a small bumpiness should be expected; if several pulsars contribute the opposite scenario is expected.   

So far no positive detection of CRE anisotropy was reported by the Fermi-LAT collaboration, but some stringent upper limits were published.  In Di Bernardo {\it et al.} 2010 \cite{electrons_final} we showed that the pulsar scenario is still compatible with these upper limits. 
Also, no evidence of spectral bumpiness has been found so far in the $e^+ + e^-$ spectrum.  

It should be noted that several astrophysical uncertainties prevent accurate predictions of the CRE anisotropy and of the spectral bumpiness. 
For example, unknown irregularities in the local structure of the Galactic magnetic field may distort the angular distribution of the CRE flux due to a nearby pulsar. 
Furthermore, due to the stochastic nature of the $e^-$ emission of nearby SNRs,  the CRE standard component is expected to be subject to fluctuations which may produce anisotropies and spectral bumpiness which may hide those due to pulsars.    


The other possible scenario to explain the origin of the extra component is more exotic but very appealing as it invokes DM annihilihation/decay as the origin of  the $e^\pm$ extra component.
Plenty of papers were published on that subject after the release of PAMELA and Fermi-LAT results (see e.g. He 2009 \cite{DM_review_2009} for a review).
That scenario, however, present some problems. The most important are the following ones.

\begin{itemize}
\item It requires a heavy DM particle mass -- O(TeV) -- and an annihilation cross section much higher than that predicted by standard cosmology if one assumes that DM is a thermal relic. 
\item Since no excess was detected for antiprotons, the annihilation/decay channels must include {\it only leptons} (lepto-philic DM). 
\end{itemize}

Although several DM models which may fulfil those conditions were developed, another issue arises when electroweak corrections are taken into account. 
Those corrections, in fact, give rise -- even in a lepto-philic scenario -- to soft electroweak gauge bosons, and hence to antiprotons, at the end of their decay chains (Ciafaloni {\it et al.} 2011 \cite{Ciafaloni_2011}).
Since those exotic ${\bar p}$ are produced mainly in the Galactic Center region, the flux reaching the Earth strongly depends on the properties of CR propagation in the Galaxy. 
As we discussed in Sec. \ref{sec:low_energy}, these properties are still subject to strong uncertainties.  
It was shown in Evoli {\it et al.} 2011 \cite{Exotic_ap_2011} that, accounting for those uncertainties,
a scenario in which a heavy DM particle annihilates into muons is still compatible with the antiproton constraints.  
In the same paper it was also shown that AMS-02 is expected to constrain even more these models since its sensitivity to antiprotons will be much higher.

\section{Conclusions and future perspectives}

In this contribution we argued as recent experimental data rule out the standard scenario in which CR positrons are produced only by CR spallation onto the ISM
and showed as an empirical model which invokes an extra $e^\pm$ component fulfils all data sets. 
We also discussed several uncertainties which still prevent to infer some of the properties of CR electron and positron sources. 
We argued that at low energy those uncertainties are dominated by our poor knowledge of CR propagation (which prevent an accurate determination of the  injection spectrum of the $e^-$ standard component) while at high energy the effect of the stochastic nature of astrophysical sources prevails (which makes more difficult to decide between the astrophysical and DM origin of the extra component).  

Forthcoming experiments like AMS-02 and CALET are expected to reduce drastically the uncertainties on the propagation parameters by providing more accurate measurements of the spectra of the nuclear components of CR. Fermi-LAT and those experiments are also expected to provide more accurate measurements of the CRE spectrum and anisotropy looking for features which may give a clue of the nature of the extra component. 

\bibliographystyle{ws-procs9x6}
\bibliography{ws-pro-sample}

\begin{thebibliography}{10}

\bibitem{Adriani_Nature_2008}
O.~A. {\it et al.}~[PAMELA~collaboration], {\em Nature} {\bf 458}, 607(April
  2009).

\bibitem{Fermi_el_2009}
A.~A.~A. {\it et al.}~[Fermi~Collaboration], {\em Physical Review Letters} {\bf
  102}, p. 181101(May 2009).

\bibitem{Fermi_el_2010}
M.~A. {\it et al.}~[Fermi~Collaboration], {\em Physical Review D} {\bf 82}, p.
  092004(November 2010).

\bibitem{pamela_el_2011}
O.~A. {\it et al.}~[PAMELA~collaboration], {\em Physical Review Letters} {\bf
  106}, p. 201101(May 2011).

\bibitem{Fermi_el_interpretation}
D.~{Grasso}, S.~{Profumo}, A.~W. {Strong}, L.~{Baldini}, R.~{Bellazzini}, E.~D.
  {Bloom}, J.~{Bregeon}, G.~{di Bernardo}, D.~{Gaggero}, N.~{Giglietto},
  T.~{Kamae}, L.~{Latronico}, F.~{Longo}, M.~N. {Mazziotta}, A.~A. {Moiseev},
  A.~{Morselli}, J.~F. {Ormes}, M.~{Pesce-Rollins}, M.~{Pohl}, M.~{Razzano},
  C.~{Sgro}, G.~{Spandre} and T.~E. {Stephens}, {\em Astroparticle Physics}
  {\bf 32}, 140(September 2009).

\bibitem{electrons_final}
G.~{di Bernardo}, C.~{Evoli}, D.~{Gaggero}, D.~{Grasso}, L.~{Maccione} and
  M.~N. {Mazziotta}, {\em Astroparticle Physics} {\bf 34}, 528(February 2011).

\bibitem{Serpico2011}
P.~D. {Serpico}, {\em ArXiv e-prints} (August 2011).

\bibitem{Pamela_p_2011}
O.~A. {\it et al.}~[PAMELA~collaboration], {\em Science} {\bf 332}, p.~69(April
  2011).

\bibitem{AharonianAtoyan1995}
A.~M. {Atoyan}, F.~A. {Aharonian} and H.~J. {V{\"o}lk}, {\em Physical Review D}
  {\bf 52}, 3265(September 1995).

\bibitem{Hooper_2009}
D.~{Hooper}, P.~{Blasi} and P.~{Dario Serpico}, {\em Journal of Cosmology and
  Astroparticle Physics} {\bf 1}, p.~25(January 2009).

\bibitem{Profumo_2008}
S.~{Profumo}, {\em ArXiv e-prints} (December 2008).

\bibitem{JaffeStrong2011}
T.~R. {Jaffe}, A.~J. {Banday}, J.~P. {Leahy}, S.~{Leach} and A.~W. {Strong},
  {\em Monthly Notices of the Royal Astronomical Society} {\bf 416},
  1152(September 2011).

\bibitem{Berezinsky_book}
V.~S. {Berezinskii}, S.~V. {Bulanov}, V.~A. {Dogiel} and V.~S. {Ptuskin}, {\em
  {Astrophysics of cosmic rays}} 1990.

\bibitem{Maurin2001}
D.~{Maurin}, F.~{Donato}, R.~{Taillet} and P.~{Salati}, {\em Astrophysical
  Journal} {\bf 555}, 585(July 2001).

\bibitem{Donato2004}
F.~{Donato}, N.~{Fornengo}, D.~{Maurin}, P.~{Salati} and R.~{Taillet}, {\em
  Physical Review D} {\bf 69}, p. 063501(March 2004).

\bibitem{anstat_2010}
G.~{di Bernardo}, C.~{Evoli}, D.~{Gaggero}, D.~{Grasso} and L.~{Maccione}, {\em
  Astroparticle Physics} {\bf 34}, 274(December 2010).

\bibitem{Trotta2011}
R.~{Trotta}, G.~{J{\'o}hannesson}, I.~V. {Moskalenko}, T.~A. {Porter}, R.~{Ruiz
  de Austri} and A.~W. {Strong}, {\em Astrophysical Journal} {\bf 729}, p.
  106(March 2011).

\bibitem{Delahaye_2010}
T.~{Delahaye}, J.~{Lavalle}, R.~{Lineros}, F.~{Donato} and N.~{Fornengo}, {\em
  Astronomy and Astrophysics} {\bf 524}, p. A51(December 2010).

\bibitem{BulanovDogel1974}
S.~V. {Bulanov} and V.~A. {Dogel}, {\em Astrophysics and Space Science} {\bf
  29}, 305(August 1974).

\bibitem{DM_review_2009}
X.-G. {He}, {\em Modern Physics Letters A} {\bf 24}, 2139 (2009).

\bibitem{Ciafaloni_2011}
P.~{Ciafaloni}, D.~{Comelli}, A.~{Riotto}, F.~{Sala}, A.~{Strumia} and
  A.~{Urbano}, {\em Journal of Cosmology and Astroparticle Physics} {\bf 3},
  p.~19(March 2011).

\bibitem{Exotic_ap_2011}
C.~{Evoli}, I.~{Cholis}, D.~{Grasso}, L.~{Maccione} and P.~{Ullio}, {\em ArXiv
  e-prints} (August 2011).

\end{thebibliography}

\end{document}